\documentstyle[prd,aps,preprint,tighten,epsfig]{revtex}

\begin{document}

\draft

\title{On the Quasi-fixed Point in the Running of CP-violating
Phases of Majorana Neutrinos}
\author{{\bf Shu Luo} ~ and ~ {\bf Zhi-zhong Xing}}
\address{CCAST (World Laboratory), P.O. Box 8730, Beijing 100080,
China \\
and Institute of High Energy Physics, Chinese Academy of Sciences,
\\
P.O. Box 918, Beijing 100049, China
\footnote{Mailing address} \\
({\it Electronic address: xingzz@mail.ihep.ac.cn}) }
\maketitle

\begin{abstract}
Taking the standard parametrization of three-flavor neutrino
mixing, we carefully examine the evolution of three CP-violating
phases $(\delta, \alpha^{}_1, \alpha^{}_2)$ with energy scales in
the realistic limit $\theta^{}_{13} \rightarrow 0$. If $m^{}_3$
vanishes, we find that the one-loop renormalization-group equation
(RGE) of $\delta$ does not diverge and its running has no
quasi-fixed point. When $m^{}_3 \neq 0$ holds, we show that the
continuity condition derived by Antusch {\it et al} is always
valid, no matter whether the $\tau$-dominance approximation is
taken or not. The RGE running of $\delta$ undergoes a quasi-fixed
point determined by a nontrivial input of $\alpha^{}_2$ in the
limit $m^{}_1 \rightarrow 0$. If three neutrino masses are nearly
degenerate, it is also possible to arrive at a quasi-fixed point
in the RGE evolution of $\delta$ from the electroweak scale to the
seesaw scale or vice versa. Furthermore, the continuity condition
and the quasi-fixed point of CP-violating phases in another useful
parametrization are briefly discussed.
\end{abstract}


\newpage

\framebox{\Large\bf 1} ~ Recent solar \cite{SNO}, atmospheric
\cite{SK}, reactor \cite{KM} and accelerator \cite{K2K} neutrino
oscillation experiments have provided us with very robust evidence
that neutrinos are massive and lepton flavors are mixed. The
phenomenon of lepton flavor mixing is described by a $3\times 3$
unitary matrix $V$. A particular parametrization of $V$ has been
advocated by the Particle Data Group \cite{PDG}:
\begin{equation}
V = \left( \matrix{ c^{}_{12}c^{}_{13} & s^{}_{12}c ^{}_{13} &
s^{}_{13} e^{-i\delta} \cr -s^{}_{12}c^{}_{23}
-c^{}_{12}s^{}_{23}s^{}_{13} e^{i\delta} & c^{}_{12}c^{}_{23}
-s^{}_{12}s^{}_{23}s^{}_{13} e^{i\delta} & s^{}_{23}c^{}_{13} \cr
s^{}_{12}s^{}_{23} -c^{}_{12}c^{}_{23}s^{}_{13} e^{i\delta} &
-c^{}_{12}s^{}_{23} -s^{}_{12}c^{}_{23}s^{}_{13} e^{i\delta} &
c^{}_{23}c^{}_{13} } \right) \left ( \matrix{e^{i\alpha^{}_1/2} &
0 & 0 \cr 0 & e^{i\alpha^{}_2/2} & 0 \cr 0 & 0 & 1 \cr} \right )
\; ,
\end{equation}
where $c^{}_{ij} \equiv \cos\theta_{ij}$ and $s^{}_{ij} \equiv
\sin\theta_{ij}$ (for $ij=12,23$ and $13$). The phase parameters
$\alpha^{}_1$ and $\alpha^{}_2$ are commonly referred to as the
Majorana CP-violating phases, because they are only physical for
Majorana neutrinos and have nothing to do with CP violation in the
neutrino-neutrino and antineutrino-antineutrino oscillations. A
global analysis of current experimental data yields \cite{Vissani}
$30^\circ < \theta_{12} < 38^\circ$, $36^\circ < \theta_{23} <
54^\circ$ and $\theta_{13} < 10^\circ$ at the $99\%$ confidence
level. In addition, the neutrino mass-squared differences $\Delta
m^2_{21} \equiv m^2_2 - m^2_1 = (7.2 \cdots 8.9) \times 10^{-5}
~{\rm eV}^2$ and $\Delta m^2_{32} \equiv m^2_3 - m^2_2 = \pm (1.7
\cdots 3.3) \times 10^{-3} ~{\rm eV}^2$ have been extracted from
solar and atmospheric neutrino oscillations at the same confidence
level \cite{Vissani}. The sign of $\Delta m^2_{32}$ remains
undetermined and three CP-violating phases of $V$ are entirely
unrestricted.

Note that $\theta^{}_{13} = 0$, which may naturally arise from an
underlying flavor symmetry (e.g., $S_3$ \cite{FX96} or $A_4$
\cite{A4}), is absolutely allowed by the present experimental
data. Note also that either $m^{}_1 = 0$ or $m^{}_3 = 0$, which
can be obtained from a specific neutrino mass model (e.g., the
minimal seesaw model \cite{MSM}), is absolutely consistent with
current neutrino oscillation data. These interesting limits
deserve some careful consideration in the study of neutrino
phenomenology. For instance, the one-loop renormalization-group
equation (RGE) of $\delta$ includes the $1/\sin\theta^{}_{13}$
term which is very dangerous in the limit $\theta^{}_{13}
\rightarrow 0$ \cite{Casas}. It has been noticed by Antusch {\it
et al} \cite{Antusch} that the derivative of $\delta$ can keep
finite when $\theta^{}_{13}$ approaches zero, if $\delta$,
$\alpha^{}_1$ and $\alpha^{}_2$ satisfy a novel continuity
condition in the $\tau$-dominance approximation (in which the
small contributions of electron and muon Yukawa couplings to the
RGEs are safely neglected). It has also been noticed by us
\cite{Luo} that the RGE running of $\delta$ may undergo a
nontrivial quasi-fixed point driven by the nontrivial inputs of
$\alpha^{}_1$ and $\alpha^{}_2$ in the tri-bimaximal neutrino
mixing scenario \cite{TB} with a near mass degeneracy of three
neutrinos.

We find it desirable to examine the continuity condition obtained
by Antusch {\it et al} \cite{Antusch} without taking the
$y^2_\tau$-dominance approximation, where $y^{}_\tau$ denotes the
tau-lepton Yukawa coupling eigenvalue. The reason is simply that
the $y^2_e$ and $y^2_\mu$ contributions to the RGE of $\delta$ may
also involve the $1/\sin\theta^{}_{13}$ terms and become dangerous
in the limit $\theta^{}_{13} \rightarrow 0$. On the other hand, it
is desirable to look at possible quasi-fixed points in the RGE
running of $\delta$ by choosing more generic neutrino mixing
scenarios with vanishing (or vanishingly small) $\theta^{}_{13}$
and considering different patterns of the neutrino mass spectrum.

The main purpose of this paper is just to carry out a careful
analysis of the RGE evolution of three CP-violating phases
$(\delta, \alpha^{}_1, \alpha^{}_2)$ in the realistic limit
$\theta^{}_{13} \rightarrow 0$ from the electroweak scale
$\Lambda^{}_{\rm EW} \sim 10^2$ GeV to the typical seesaw scale
$\Lambda^{}_{\rm SS} \sim 10^{14}$ GeV. If $m^{}_3$ vanishes, we
find that the RGE of $\delta$ does not diverge and its running has
no quasi-fixed point. This new observation demonstrates that our
previous understanding of the running behaviors of $\delta$ is
more or less incomplete. When $m^{}_3 \neq 0$ holds, we show that
the continuity condition derived by Antusch {\it et al} can be
rediscovered even though the $y^2_e$ and $y^2_\mu$ contributions
to the RGE of $\delta$ are not neglected. The RGE running of
$\delta$ undergoes a quasi-fixed point determined by a nontrivial
input of $\alpha^{}_2$ in the limit $m^{}_1 \rightarrow 0$. If
three neutrino masses are nearly degenerate (either $\Delta
m^2_{32} >0$ or $\Delta m^2_{32} < 0$), a quasi-fixed point may
also show up in the RGE evolution of $\delta$ from the electroweak
scale to the seesaw scale (or vice versa). Finally we give some
brief comments on the continuity condition and the quasi-fixed
point of CP-violating phases in another useful parametrization of
$V$.

\vspace{0.5cm}

\framebox{\Large\bf 2} ~ The exact one-loop RGEs of three neutrino
masses $(m^{}_1, m^{}_2, m^{}_3)$, three mixing angles
$(\theta^{}_{12}, \theta^{}_{23}, \theta^{}_{13})$ and three
CP-violating phases $(\delta, \alpha^{}_1, \alpha^{}_2)$ have
already been derived by Antusch {\it et al} \cite{Antusch} and can
be found from the web page \cite{WP}
\footnote{Note that the Majorana phases $\varphi^{}_1$ and
$\varphi^{}_2$ defined in Refs. \cite{Antusch} and \cite{WP} are
equivalent to $\alpha^{}_1$ and $\alpha^{}_2$ defined in Eq. (1):
$\varphi^{}_i = - \alpha^{}_i$ (for $i=1,2$).}.
Their results, which have been confirmed by Mei and Zhang
independently \cite{Mei}, clearly show that only the RGE of
$\delta$ contains the $1/\sin\theta^{}_{13}$ term. For simplicity,
here we only write out the derivative of $\delta$ in an exact but
compact way:
\begin{equation}
\frac{{\rm d}\delta}{{\rm d}t} \; =\; \frac{C \left (y^2_\tau -
y^2_\mu \right )}{32 \pi^2} \cdot \frac{m^{}_3 \chi}{\Delta
m^2_{31} \Delta m^2_{32}} \cdot \frac{\sin 2\theta^{}_{12} \sin
2\theta^{}_{23}}{\sin\theta^{}_{13}} ~ + ~ {\rm other ~ terms} \;
,
\end{equation}
where $t\equiv \ln (\mu/\Lambda_{\rm SS})$ with $\mu$ being an
arbitrary renormalization scale below $\Lambda_{\rm SS}$ but above
$\Lambda_{\rm EW}$, $C=-3/2$ in the standard model (SM) or $C=1$
in the minimal supersymmetric standard model (MSSM),
\begin{equation} \chi \; =\; m^{}_3 \Delta m^2_{21} \sin\delta +
m^{}_2 \Delta m^2_{31} \sin \left (\delta + \alpha^{}_2 \right ) -
m^{}_1 \Delta m^2_{32} \sin \left (\delta + \alpha^{}_1 \right )
\; ,
\end{equation}
and ``other terms" stand for those terms which do not include the
$1/\sin\theta^{}_{13}$ factor. We find that the $y^2_e$
contribution to ${\rm d}\delta/{\rm d}t$ does not involve
$1/\sin\theta^{}_{13}$ at all, while the $1/\sin\theta^{}_{13}$
terms associated with $y^2_\mu$ and $y^2_\tau$ contributions to
${\rm d}\delta/{\rm d}t$ are identical in magnitude but have the
opposite sign. When the $\tau$-dominance approximation is taken
(i.e., neglecting the $y^2_e$ and $y^2_\mu$ contributions in the
RGEs), Eq. (2) reproduces the approximate $1/\sin\theta^{}_{13}$
term of ${\rm d}\delta/{\rm d}t$ given in Ref. \cite{Antusch}.

In the limit $\theta^{}_{13} \rightarrow 0$, which is allowed (and
even favored \cite{Vissani}) by current neutrino oscillation data,
the $1/\sin\theta^{}_{13}$ term in ${\rm d}\delta/{\rm d}t$
diverges. To keep ${\rm d}\delta/{\rm d}t$ finite, the divergence
of $1/\sin\theta^{}_{13}$ has to be cancelled by its associate
factor. Eq. (2) indicates that $m^{}_3 \chi = 0$ needs to be
satisfied, in order to cancel the divergence induced by
$1/\sin\theta^{}_{13}$ in the limit $\theta^{}_{13} \rightarrow
0$. There are two separate possibilities:
\begin{enumerate}
\item    $m^{}_3 = 0$. This special but interesting possibility
was {\it not} mentioned in Ref. \cite{Antusch}. In this case, the
derivative of $\delta$ is apparently finite for vanishing or
vanishingly small $\theta^{}_{13}$. Hence the corresponding RGE
running of $\delta$ is expected to be mild and have no quasi-fixed
point. Note that only the difference between $\alpha^{}_1$ and
$\alpha^{}_2$ has physical significance in the limit $m^{}_3
\rightarrow 0$, just like the instructive case in the minimal
seesaw model with two heavy right-handed Majorana neutrinos
\cite{Guo}. If both $m^{}_3 = 0$ and $\theta^{}_{13} =0$ hold at a
given energy scale, one can easily show that $m^{}_3$ and
$\theta^{}_{13}$ will keep vanishing at any scale between
$\Lambda^{}_{\rm EW}$ and $\Lambda^{}_{\rm SS}$ \cite{Grimus}
\footnote{Note that the one-loop RGEs of $m^{}_i$ (for $i=1,2,3$)
have the form ${\rm d}m^{}_i/{\rm d}t \propto m^{}_i$ \cite{WP}.
Hence $m^{}_i =0$ keeps unchanged if it is initially given at one
energy scale.}.
In this particular case, $\delta$ is not well-defined and has no
physical meaning at all energy scales.

\item    $\chi =0$. In this case, one may arrive at the continuity
condition from Eq. (3):
\begin{equation}
\cot\delta \; =\; \frac{ m^{}_1 \cos\alpha^{}_1 - \left ( 1 +
\zeta \right ) m^{}_2 \cos\alpha^{}_2 - \zeta m^{}_3 } {\left ( 1
+ \zeta \right ) m^{}_2 \sin\alpha^{}_2 - m^{}_1 \sin\alpha^{}_1}
\; ,
\end{equation}
where $\zeta \equiv \Delta m^2_{21}/\Delta m^2_{32} \approx \pm
(2.2\cdots 5.2) \times 10^{-2}$. Although this result is
equivalent to the one derived by Antusch {\it et al} in the
$\tau$-dominance approximation \cite{Antusch}, it is now obtained
by us in no special assumption or approximation. Given
$\alpha^{}_1 = \alpha^{}_2 = 0$, the resultant quasi-fixed point
is trivially $\delta = 0$ or $\delta = \pi$. A nontrivial
quasi-fixed point in the RGE running of $\delta$ can in general
result from the nontrivial inputs of $\alpha^{}_1$ and
$\alpha^{}_2$.
\end{enumerate}
In the following, we shall take some typical examples to
illustrate the quasi-fixed point in the RGE evolution of $\delta$
from $\Lambda^{}_{\rm EW}$ to $\Lambda^{}_{\rm SS}$ either in the
SM or in the MSSM.

\vspace{0.5cm}

\framebox{\Large\bf 3} ~ In view of the fact that the absolute
mass scale of three light neutrinos and the sign of $\Delta
m^2_{32}$ remain unknown, let us consider four possible patterns
of the neutrino mass spectrum to simplify the continuity condition
in Eq. (4) and discuss the quasi-fixed point in the RGE running of
$\delta$ by taking a few typical numerical examples
\footnote{Our numerical calculations follow a ``running and
diagonalizing" procedure \cite{Antusch}: we first compute the RGE
evolution of lepton mass matrices starting from $\Lambda^{}_{\rm
EW}$, and then extract their mass eigenvalues and flavor mixing
parameters at $\Lambda_{\rm SS}$. This approach itself is
independent of our analytical derivation of the RGEs for three
neutrino masses $(m^{}_1, m^{}_2, m^{}_3)$, three mixing angles
$(\theta^{}_{12}, \theta^{}_{23}, \theta^{}_{13})$ and three
CP-violating phases $(\delta, \alpha^{}_1, \alpha^{}_2)$. If the
limit $\theta^{}_{13} \rightarrow 0$ is taken at $\Lambda_{\rm
EW}$, any initial input of $\delta$ is allowed but it does not
take any effect in the RGE running. The finite running result of
$\delta$ is actually attributed to the initial values of two
Majorana phases $\alpha^{}_1$ and $\alpha^{}_2$.}.

{\bf (1) Normal hierarchy:} $m^{}_1 \ll m^{}_2 \ll m^{}_3$. For
simplicity, we typically take $m^{}_1 = 0$ at $\Lambda^{}_{\rm
EW}$ in our analysis. Then $m^{}_2=\sqrt{\Delta m^2_{21}} \approx
8.9 \times 10^{-3}$ eV and $m^{}_3 = \sqrt{|\Delta m^2_{32}| +
\Delta m^2_{21}} \approx 5.1 \times 10^{-2}$ eV can be obtained
from the best-fit values $\Delta m^2_{21} = 8.0 \times 10^{-5}$ eV
and $|\Delta m^2_{32}| = 2.5 \times 10^{-3}$ eV \cite{Vissani}. In
this special case, the phase parameter $\alpha^{}_1$ has no
physical significance and Eq. (4) can easily be simplified to
\begin{equation}
m^{}_2 \sin \delta + m^{}_3 \sin \left ( \delta + \alpha^{}_2
\right ) \; = \; 0 \; .
\end{equation}
Once the initial value of $\alpha^{}_2$ is fixed at
$\Lambda^{}_{\rm EW}$, the value of $\delta$ at its quasi-fixed
point can be determined.

To illustrate, we present our typical numerical examples in Table
I. Because $m^{}_1 =0$ and $\theta^{}_{13} =0$ are taken at
$\Lambda^{}_{\rm EW}$, the corresponding phase parameters
$\alpha^{}_1$ and $\delta$ are not well-defined and their initial
values at $\Lambda^{}_{\rm EW}$ have no physical meaning. While
$m^{}_1 =0$ keeps unchanged up to the seesaw scale,
$\theta^{}_{13}$ will become non-vanishing due to the radiative
correction. It is also the radiative correction that gives rise to
nontrivial values of $\delta$ at $\mu
> \Lambda^{}_{\rm EW}$, but $\alpha^{}_1$ remains arbitrary and
has no physical significance as a direct consequence of $m^{}_1
=0$ from $\Lambda^{}_{\rm EW}$ to $\Lambda^{}_{\rm SS}$. Of
course, $\delta$ and $\alpha^{}_2$ must satisfy Eq. (5) at the
quasi-fixed point of $\delta$.

Table I indicates that the numerical changes of three mixing
angles from $\Lambda^{}_{\rm EW}$ to $\Lambda^{}_{\rm SS}$ are
negligibly small. This result is generally expected for the normal
mass hierarchy of three light neutrinos \cite{RGE}. The radiative
correction to the Majorana phase $\alpha^{}_2$ is also vanishingly
small, as one can see from Table I. The nontrivial output of
$\delta$ comes from the nontrivial input of $\alpha^{}_2$ via the
RGE running effects. Of course, larger values of $\tan\beta$ will
in general lead to larger radiative corrections to the neutrino
mass and mixing parameters in the MSSM.

{\bf (2) Inverted hierarchy:} $m^{}_3 \ll m^{}_1 \lesssim m^{}_2
$. We have pointed out that the derivative of $\delta$ must be
finite when $m^{}_3 =0$ holds; i.e., there is no quasi-fixed point
in the RGE running of $\delta$ in the limit $m^{}_3 \rightarrow
0$. If $m^{}_3$ is very small but non-vanishing, however, the
evolution of $\delta$ may undergo a quasi-fixed point in the limit
$\theta^{}_{13} \rightarrow 0$. For illustration, we typically
take $m^{}_3= 10^{-5}$ eV at $\Lambda^{}_{\rm EW}$ in our
analysis. Then we obtain $m^{}_2 = \sqrt{|\Delta m^2_{32}| +
m^2_3} \approx 5.0 \times 10^{-2}$ eV and $m^{}_1 \approx
\sqrt{|\Delta m^2_{32}| - \Delta m^2_{21} + m^2_3} \approx 4.9
\times 10^{-2}$ eV from the best-fit values $\Delta m^2_{21} = 8.0
\times 10^{-5}$ eV and $|\Delta m^2_{32}| = 2.5 \times 10^{-3}$ eV
\cite{Vissani}. The first term on the right-hand side of Eq. (3)
is safely negligible. As a result, $\chi =0$ leads to an
approximate relation for three phase parameters at the quasi-fixed
point of $\delta$:
\begin{equation}
m^{}_1 \sin \left ( \delta + \alpha^{}_2 \right ) \; \approx \;
m^{}_2 \sin \left ( \delta + \alpha^{}_1 \right ) \; .
\end{equation}
The value of $\delta$ can therefore be determined from this
equation together with the initial inputs of $\alpha^{}_1$ and
$\alpha^{}_2$ at $\Lambda^{}_{\rm EW}$.

We present a few numerical examples in Table II for illustration.
Although $\theta^{}_{13} =0$ is taken at $\Lambda^{}_{\rm EW}$, it
will become non-vanishing at $\mu > \Lambda^{}_{\rm EW}$ due to
the radiative correction. The phase parameter $\delta$, whose RGE
running is driven by the initial values of two Majorana phases
$\alpha^{}_1$ and $\alpha^{}_2$, turns out to be nontrivial at
$\mu > \Lambda^{}_{\rm EW}$ and keeps on staying at its
quasi-fixed point. It is easy to check that Eq. (6) is well
satisfied by the values of three CP-violating phases at
$\Lambda^{}_{\rm SS}$ as shown in Table II. A particularly
interesting possibility is $\alpha^{}_1 \approx \alpha^{}_2$ (Case
III in Table II). In this case, the sum of $\delta$ and
$\alpha^{}_1$ (or $\alpha^{}_2$) is close to $n\pi$ (for $n=0,
\pm1, \pm 2, \cdots$), which can also be seen from Eq. (6).

{\bf (3) Near degeneracy with $\Delta m^2_{32} >0$:} $m^{}_1
\lesssim m^{}_2 \lesssim m^{}_3$. For simplicity, we typically
take $m^{}_1 = 0.2~{\rm eV}$ at $\Lambda^{}_{\rm EW}$ in our
numerical calculations. Then $m^{}_1 \approx m^{}_2 \approx
m^{}_3$ automatically holds, and the continuity condition $\chi
=0$ for the RGE running of $\delta$ in the limit $\theta^{}_{13}
\rightarrow 0$ can approximately be simplified to
\begin{equation}
\Delta m^{2}_{32} \left [ \sin \left ( \delta + \alpha^{}_1 \right
) - \sin \left ( \delta + \alpha^{}_2 \right ) \right ] \; \approx
\; \Delta m^{2}_{21} \left [ \sin \delta + \sin \left ( \delta +
\alpha^{}_2 \right ) \right ] \; .
\end{equation}
This equation may further be simplified in two different cases:
\begin{itemize}
\item     If $\alpha^{}_1 \neq \alpha^{}_2$ holds, the right-hand
side of Eq. (7) is strongly suppressed by the smallness of $\Delta
m^2_{21}$, leading to
\begin{equation}
\sin \left (\delta + \alpha^{}_1 \right ) \; \approx \; \sin \left
(\delta + \alpha^{}_2 \right ) \; ;
\end{equation}
namely, $\delta \approx -(\alpha^{}_1 + \alpha^{}_2)/2 + (n
+1/2)\pi$ (for $n = 0, \pm 1, \pm 2, \cdots$) can be achieved.

\item     If $\alpha^{}_1 \approx \alpha^{}_2$ holds, the
left-hand side of Eq. (7) is either vanishing or vanishingly
small, leading to
\begin{equation}
\sin\delta + \sin \left (\delta + \alpha^{}_2 \right ) \; \approx
\; 0 \; ;
\end{equation}
namely, $\delta \approx -\alpha^{}_2/2 + n\pi$ (for $n = 0, \pm 1,
\pm 2, \cdots$) can be obtained.
\end{itemize}
These two simple relations have been found by us in Ref.
\cite{Luo}, where radiative corrections to the generalized
tri-bimaximal neutrino mixing pattern is discussed in detail from
the seesaw scale down to the electroweak scale.

A few numerical examples are given in Table III for illustration.
One can see that $\delta = 0$ at $\mu > \Lambda^{}_{\rm EW}$ is a
straightforward consequence of $\alpha^{}_1 = \alpha^{}_2 =0$ at
$\mu = \Lambda^{}_{\rm EW}$ (Case I), implying that CP keeps to be
a good symmetry in the RGE evolution. This result illustrates that
the nontrivial running of $\delta$ must be driven by the
nontrivial inputs of $\alpha^{}_1$ and $\alpha^{}_2$, if the limit
$\theta^{}_{13} \rightarrow 0$ is taken at $\Lambda^{}_{\rm EW}$.
Note that the outputs of $\delta$, $\alpha^{}_1$ and $\alpha^{}_2$
in Case II of Table III, where the inputs of $\alpha^{}_1$ and
$\alpha^{}_2$ are different from each other, satisfy the
analytical approximation in Eq. (8) very well. On the other hand,
the numerical example shown in Case III of Table III ($\alpha^{}_1
= \alpha^{}_2$ at $\Lambda^{}_{\rm EW}$) is consistent with the
analytical approximation in Eq. (9).

{\bf (4) Near degeneracy with $\Delta m^2_{32} <0$:} $m^{}_3
\lesssim m^{}_1 \lesssim m^{}_2$. For simplicity, we typically
take $m^{}_1 = 0.2~{\rm eV}$ at $\Lambda^{}_{\rm EW}$ in our
analysis. We find that the analytical approximations obtained in
Eqs. (7), (8) and (9) for the continuity condition of ${\rm
d}\delta/{\rm d}t$ in the limit $\theta^{}_{13} \rightarrow 0$ are
also applicable for the case of $\Delta m^2_{32} <0$. Three
numerical examples are presented in Table IV to illustrate the RGE
evolution of nine neutrino mixing parameters. Comparing between
Tables III and IV, one can see that the sign flip of $\Delta
m^2_{32}$ only causes very mild changes of the RGE running
behaviors of relevant quantities. When $\alpha^{}_1 \neq
\alpha^{}_2$ holds (Case II of Tables III and IV), flipping the
sign of $\Delta m^2_{32}$ is simply equivalent to shifting the
output value of $\delta$ by $\pi$. The latter is certainly
compatible with the solution to Eq. (8), which has the ambiguities
of $n\pi$ (for $n=0, \pm 1,\pm 2, \cdots$). When $\alpha^{}_1
\approx \alpha^{}_2$ holds (Case III of Tables III and IV), the
quasi-fixed point of $\delta$ is almost insensitive to the sign
flip of $\Delta m^2_{32}$.

Because three neutrino masses are nearly degenerate, the relevant
mixing angles and CP-violating phases may get significant
radiative corrections \cite{RGE}. This well-known feature can be
seen either for $\Delta m^2_{32} > 0$ or for $\Delta m^2_{32} < 0$
and either in the SM or in the MSSM.

\vspace{0.5cm}

\framebox{\Large\bf 4} ~ We have examined possible quasi-fixed
points in the RGE evolution of three leptonic CP-violating phases
$(\delta, \alpha^{}_1, \alpha^{}_2)$ by taking the realistic limit
$\theta^{}_{13} \rightarrow 0$ in the standard parametrization of
three-flavor neutrino mixing. While the Majorana phases
$\alpha^{}_1$ and $\alpha^{}_2$ do not undergo any divergence in
their RGE running from the electroweak scale to the seesaw scale,
the phase parameter $\delta$ may in general have a quasi-fixed
point. This interesting point has essentially been observed in
Refs. \cite{Antusch} and \cite{Luo}, where the RGE of $\delta$ is
derived in the $\tau$-dominance approximation. Our present
analysis, which is based on the exact RGE of $\delta$, has more
generally demonstrated that there may only exist a single
nontrivial quasi-fixed point and it exactly obeys the continuity
condition obtained before. We have also noticed that the
derivative of $\delta$ will not diverge and its running has no
quasi-fixed point, if $m^{}_3$ vanishes. This new observation
allows us to have a complete insight into the correlation between
the neutrino mass spectrum and the RGE running of $\delta$. Taking
four typical patterns of the neutrino mass spectrum, we have
explicitly illustrated how the evolution of $\delta$ undergoes a
quasi-fixed point determined by the nontrivial inputs of
$\alpha^{}_1$ and $\alpha^{}_2$ in the limit $\theta^{}_{13}
\rightarrow 0$.

Let us remark that our physical understanding of the quasi-fixed
point(s) in the RGE evolution of leptonic CP-violating phases is
not subject to any concrete parametrization of the neutrino mixing
matrix. Nevertheless, a good parametrization may simplify our
analytical calculations and make the underlying physics more
transparent \cite{FX}. As recently pointed out in Ref.
\cite{Xing06}, the one-loop RGEs of neutrino mixing parameters
will take very simple and instructive forms in the
$\tau$-dominance approximation, if the lepton mixing matrix $V$ is
parametrized as
\footnote{Note that the phase convention of $V$ in Eq. (10) is
slightly different from that taken in Ref. \cite{Xing06}. The
present choice allows the phase factor $e^{\pm i\phi}$ to
automatically vanish in the limit $\theta^{}_l \rightarrow 0$,
thus it is more convenient for a numerical analysis of the
quasi-fixed point in the RGE running of $\phi$. To leading order,
we have $\theta^{}_{12} \approx \theta^{}_{\nu}$, $\theta^{}_{23}
\approx \theta$, $\theta^{}_{13} \approx \theta^{}_{l}
\sin\theta$, $\delta \approx \phi$, $\alpha^{}_1 \approx 2\rho$
and $\alpha^{}_2 \approx 2\sigma$ for the relations of two
parametrizations.}:
\begin{eqnarray}
V & = & \left( \matrix{ c^{}_{l} & s^{}_{l} e^{-i\phi} & 0 \cr -
s^{}_{l} e^{+i\phi} & c^{}_{l} & 0 \cr 0 & 0 & 1 } \right) \left(
\matrix{ 1 & 0 & 0 \cr 0 & c & s \cr 0 & - s & c } \right) \left(
\matrix{ c^{}_{\nu} & s^{}_{\nu} & 0 \cr - s^{}_{\nu} & c^{}_{\nu}
& 0 \cr 0 & 0 & 1 } \right) \left ( \matrix{e^{i\rho} & 0 & 0 \cr
0 & e^{i\sigma} & 0 \cr 0 & 0 & 1 \cr} \right ) \;
\nonumber \\
& = & \left( \matrix{ c^{}_{l} c^{}_{\nu} - s^{}_{l} s^{}_{\nu} c
e^{-i\phi} & c^{}_{l} s^{}_{\nu} - s^{}_{l} c^{}_{\nu} c
e^{-i\phi} & s^{}_{l} s e^{-i\phi} \cr - c^{}_{l} s^{}_{\nu} c -
s^{}_{l} c^{}_{\nu} e^{+i\phi} & c^{}_{l} c^{}_{\nu} c - s^{}_{l}
s^{}_{\nu} e^{+i\phi} & c^{}_{l} c \cr s^{}_{\nu} s & - c^{}_{\nu}
s & c } \right) \left ( \matrix{e^{i\rho} & 0 & 0 \cr 0 &
e^{i\sigma} & 0 \cr 0 & 0 & 1 \cr} \right ) \; ,
\end{eqnarray}
where $c^{}_{l} \equiv \cos\theta^{}_{l}$, $s^{}_{l} \equiv
\sin\theta^{}_{l}$, $c^{}_{\nu} \equiv \cos\theta^{}_{\nu}$,
$s^{}_{\nu} \equiv \sin\theta^{}_{\nu}$, $c \equiv \cos\theta$ and
$s \equiv \sin\theta$. In order to keep the RGE of $\phi$ finite
in the realistic limit $\theta^{}_{l} \rightarrow 0$ (equivalent
to $\theta^{}_{13} \rightarrow 0$ in the standard
parametrization), three CP-violating phases must satisfy the
continuity condition
\begin{equation}
\frac{\left (m^2_1 + m^2_3 \right ) \sin\phi + 2 m^{}_1 m^{}_3
\sin \left (\phi + 2\rho \right )}{\Delta m^2_{31}} \; =\;
\frac{\left (m^2_2 + m^2_3 \right ) \sin\phi + 2 m^{}_2 m^{}_3
\sin \left (\phi + 2\sigma \right )}{\Delta m^2_{32}} \; .
\end{equation}
It is straightforward to check that Eq. (11) holds trivially for
arbitrary values of $\phi$, $\rho$ and $\sigma$ in the limit
$m^{}_3 \rightarrow 0$, implying that there must be no nontrivial
fixed point in the evolution of $\phi$ from the electroweak scale
to the seesaw scale (or vice versa). This continuity condition is
therefore equivalent to $m^{}_3 \chi =0$ (namely, $m^{}_3 =0$ or
$\chi =0$) which can be extracted from Eq. (2) in the standard
parametrization. But similar to $\alpha^{}_1$ and $\alpha^{}_2$,
the Majorana phases $\rho$ and $\sigma$ do not undergo any
quasi-fixed point in their RGE running.

We conclude that radiative corrections to a specific neutrino mass
model have to be taken into account in a very careful way, in
particular when there exists the quasi-fixed point in the running
of its phase parameters. To test a theoretical model or a
phenomenological ansatz, it is crucial to measure the smallest
mixing angle $\theta^{}_{13}$ (or $\theta^{}_l$) and the
CP-violating phase $\delta$ (or $\phi$) in the future neutrino
oscillation experiments. Any experimental information about the
Majorana phases $\alpha^{}_1$ and $\alpha^{}_2$ (or $\rho$ and
$\sigma$) is extremely useful in order to distinguish one model
from another, e.g., by looking at their different sensitivities to
radiative corrections.

\vspace{0.5cm}

{\it Acknowledgments:} ~ One of the authors (Z.Z.X.) is indebted
to X.D. Ji for his warm hospitality at University of Maryland,
where this paper was finalized. He is also grateful to M. Lindner
for useful discussions during the ISS neutrino workshop at Boston
University, where this paper was written, and to J.R. Espinosa for
useful comments via e-mail. Both of the authors would like to
thank H. Zhang for many helpful interactions in dealing with the
exact RGEs. Our research was supported in part by the National
Natural Science Foundation of China.

\newpage

\begin{table}
\caption{{\bf Normal hierarchy} with $m^{}_1 =0$: numerical
examples for radiative corrections to neutrino masses, lepton
flavor mixing angles and CP-violating phases from $\Lambda_{\rm
EW} \sim 10^2$ GeV to $\Lambda_{\rm SS} \sim 10^{14}$ GeV. The
Higgs mass $m^{}_H = 140$ GeV (SM) or $\tan\beta = 10$ (MSSM) has
typically been input in our calculation.}
\begin{center}
\begin{tabular}{l|ll|ll|ll}
SM & \multicolumn{2}{c}{Case I} & \multicolumn{2}{c}{Case II}
& \multicolumn{2}{c}{Case III}  \\
& $\Lambda_{\rm EW}$ & $\Lambda_{\rm SS}$ & $\Lambda_{\rm EW}$ &
$\Lambda_{\rm SS}$
& $\Lambda_{\rm EW}$ & $\Lambda_{\rm SS}$ \\
\hline
$m^{}_1 ~({\rm eV} )$ & 0 & 0 & 0 & 0 & 0 & 0 \\
$\Delta m^2_{21} ~( 10^{-5} ~{\rm eV}^2 )$ & 8.0 & 15.63 & 8.0
& 15.63 & 8.0 & 15.63 \\
$\Delta m^2_{32} ~( 10^{-3} ~{\rm eV}^2 )$ & 2.5 & 4.89 & 2.5
& 4.89 & 2.5 & 4.89 \\
\hline
$\theta_{12}$ ($^\circ$) & 34 & 34.00 & 34
& 34.00 & 34 & 34.00 \\
$\theta_{23}$ ($^\circ$) & 45 & 45.00 & 45
& 45.00 & 45 & 45.00 \\
$\theta_{13}$ ($^\circ$) & 0 & $1.59 \times 10^{-4}$ & 0 &
$1.51 \times 10^{-4}$ & 0 & $1.40 \times 10^{-4}$ \\
\hline
$\delta$ ($^\circ$) & -- & 334.37 & -- & 307.98
& -- & 279.99 \\
$\alpha^{}_1$ ($^\circ$) & -- & -- & -- & --
& -- & -- \\
$\alpha^{}_2$ ($^\circ$) & 30 & 30.00 & 60 & 60.00 & 90 & 90.00 \\
\hline\hline
MSSM & \multicolumn{2}{c}{Case I} & \multicolumn{2}{c}{Case II}
& \multicolumn{2}{c}{Case III}  \\
& $\Lambda_{\rm EW}$ & $\Lambda_{\rm SS}$ & $\Lambda_{\rm EW}$ &
$\Lambda_{\rm SS}$
& $\Lambda_{\rm EW}$ & $\Lambda_{\rm SS}$ \\
\hline
$m^{}_1 ~({\rm eV} )$ & 0 & 0 & 0 & 0 & 0 & 0 \\
$\Delta m^2_{21} ~( 10^{-5} ~{\rm eV}^2 )$ & 8.0 & 11.59
& 8.0 & 11.59 & 8.0 & 11.59 \\
$\Delta m^2_{32} ~( 10^{-3} ~{\rm eV}^2 )$ & 2.5 & 3.63
& 2.5 & 3.63 & 2.5 & 3.63 \\
\hline
$\theta_{12}$ ($^\circ$) & 34 & 33.98 & 34
& 33.98 & 34 & 33.98 \\
$\theta_{23}$ ($^\circ$) & 45 & 44.95 & 45
& 44.95 & 45 & 44.96 \\
$\theta_{13}$ ($^\circ$) & 0 & $7.51 \times 10^{-3}$ & 0 &
$7.14 \times 10^{-3}$ & 0 & $6.60 \times 10^{-3}$ \\
\hline
$\delta$ ($^\circ$) & -- & 154.37 & -- & 127.98
& -- & 99.98 \\
$\alpha^{}_1$ ($^\circ$) & -- & -- & -- & --
& -- & -- \\
$\alpha^{}_2$ ($^\circ$) & 30 & 30.00 & 60 & 60.00
& 90 & 90.00 \\
\end{tabular}
\end{center}
\end{table}

\begin{table}
\caption{{\bf Inverted hierarchy} with $m^{}_3 = 10^{-5}$ eV:
numerical examples for radiative corrections to neutrino masses,
lepton flavor mixing angles and CP-violating phases from
$\Lambda_{\rm EW} \sim 10^2$ GeV to $\Lambda_{\rm SS} \sim
10^{14}$ GeV. The Higgs mass $m^{}_H = 140$ GeV (SM) or $\tan\beta
= 10$ (MSSM) has typically been input in our calculation.}
\begin{center}
\begin{tabular}{l|ll|ll|ll}
SM & \multicolumn{2}{c}{Case I} & \multicolumn{2}{c}{Case II}
& \multicolumn{2}{c}{Case III}  \\
& $\Lambda_{\rm EW}$ & $\Lambda_{\rm SS}$ & $\Lambda_{\rm EW}$ &
$\Lambda_{\rm SS}$
& $\Lambda_{\rm EW}$ & $\Lambda_{\rm SS}$ \\
\hline
$m^{}_3 ~({\rm eV} )$ & $10^{-5}$ & $1.40 \times 10^{-5}$
& $10^{-5}$ & $1.40 \times 10^{-5}$ & $10^{-5}$ & $1.40 \times 10^{-5}$ \\
$\Delta m^2_{21} ~( 10^{-5} ~{\rm eV}^2 )$ & 8.0 & 15.63
& 8.0 & 15.63 & 8.0 & 15.63 \\
$\Delta m^2_{32} ~( 10^{-3} ~{\rm eV}^2 )$ & $-2.5$ & $-4.89$
& $-2.5$ & $-4.89$ & $-2.5$ & $-4.89$ \\
\hline
$\theta_{12}$ ($^\circ$) & 34 & 34.04 & 34 & 34.04 & 34 & 34.05 \\
$\theta_{23}$ ($^\circ$) & 45 & 45.00 & 45 & 45.00 & 45 & 45.00 \\
$\theta_{13}$ ($^\circ$) & 0 & $1.53 \times 10^{-7}$ & 0 &
$7.90 \times 10^{-8}$ & 0 & $2.48 \times 10^{-9}$ \\
\hline
$\delta$ ($^\circ$) & -- & 240.85
& -- & 241.77 & -- & 330.01 \\
$\alpha^{}_1$ ($^\circ$) & 60 & 59.94
& 45 & 44.97 & 30 & 30.00 \\
$\alpha^{}_2$ ($^\circ$) & 0 & $-0.03$ & 15 & 14.98 & 30 & 30.00
\\ \hline\hline
MSSM & \multicolumn{2}{c}{Case I} & \multicolumn{2}{c}{Case II}
& \multicolumn{2}{c}{Case III} \\
& $\Lambda_{\rm EW}$ & $\Lambda_{\rm SS}$ & $\Lambda_{\rm EW}$ &
$\Lambda_{\rm SS}$
& $\Lambda_{\rm EW}$ & $\Lambda_{\rm SS}$ \\
\hline
$m^{}_3 ~({\rm eV} )$ & $10^{-5}$ & $1.20 \times 10^{-5}$
& $10^{-5}$ & $1.20 \times 10^{-5}$ & $10^{-5}$ & $1.20 \times 10^{-5}$ \\
$\Delta m^2_{21} ~( 10^{-5} ~{\rm eV}^2 )$ & 8.0 & 11.97
& 8.0 & 11.98 & 8.0 & 11.98 \\
$\Delta m^2_{32} ~( 10^{-3} ~{\rm eV}^2 )$ & $-2.5$ & $-3.62$
& $-2.5$ & $-3.62$ & $-2.5$ & $-3.62$ \\
\hline
$\theta_{12}$ ($^\circ$) & 34 & 32.40 & 34 & 32.00 & 34 & 31.85 \\
$\theta_{23}$ ($^\circ$) & 45 & 45.04 & 45 & 45.04 & 45 & 45.04 \\
$\theta_{13}$ ($^\circ$) & 0 & $7.20 \times 10^{-6}$ & 0 &
$3.73 \times 10^{-6}$ & 0 & $1.17 \times 10^{-7}$ \\
\hline
$\delta$ ($^\circ$) & -- & 58.71
& -- & 60.48 & -- & 149.94 \\
$\alpha^{}_1$ ($^\circ$) & 60 & 62.87
& 45 & 46.68 & 30 & 30.00 \\
$\alpha^{}_2$ ($^\circ$) & 0 & 1.23
& 15 & 15.71 & 30 & 30.00 \\
\end{tabular}
\end{center}
\end{table}

\begin{table}
\caption{{\bf Near degeneracy} with $\Delta m^2_{32} > 0$:
numerical examples for radiative corrections to neutrino masses,
lepton flavor mixing angles and CP-violating phases from
$\Lambda_{\rm EW} \sim 10^2$ GeV to $\Lambda_{\rm SS} \sim
10^{14}$ GeV. The Higgs mass $m^{}_H = 140$ GeV (SM) or $\tan\beta
= 10$ (MSSM) has typically been input in our calculation.}
\begin{center}
\begin{tabular}{l|ll|ll|ll}
SM & \multicolumn{2}{c}{Case I} & \multicolumn{2}{c}{Case II }
& \multicolumn{2}{c}{Case III}  \\
& $\Lambda_{\rm EW}$ & $\Lambda_{\rm SS}$ & $\Lambda_{\rm EW}$ &
$\Lambda_{\rm SS}$
& $\Lambda_{\rm EW}$ & $\Lambda_{\rm SS}$ \\
\hline
$m^{}_1 ~({\rm eV} )$ & 0.2 & 0.28 & 0.2 & 0.28 & 0.2 & 0.28 \\
$\Delta m^2_{21} ~( 10^{-5} ~{\rm eV}^2 )$ & 8.0 & 15.47 & 8.0
& 15.47 & 8.0 & 15.47 \\
$\Delta m^2_{32} ~( 10^{-3} ~{\rm eV}^2 )$ & 2.5 & 4.88 & 2.5
& 4.89 & 2.5 & 4.88 \\
\hline
$\theta_{12}$ ($^\circ$) & 34 & 34.77 & 34
& 34.02 & 34 & 34.77 \\
$\theta_{23}$ ($^\circ$) & 45 & 45.05 & 45
& 45.02 & 45 & 45.03 \\
$\theta_{13}$ ($^\circ$) & 0 & $8.00 \times 10^{-4}$ & 0 &
$2.43 \times 10^{-2}$ & 0 & $6.13 \times 10^{-4}$ \\
\hline
$\delta$ ($^\circ$) & -- & 0 & -- & 271.06
& -- & 319.99 \\
$\alpha^{}_1$ ($^\circ$) & 0 & 0 & 260 & 259.61
& 80 & 80.00 \\
$\alpha^{}_2$ ($^\circ$) & 0 & 0 & 100 & 99.82 & 80 & 80.00 \\
\hline\hline
MSSM & \multicolumn{2}{c}{Case I} & \multicolumn{2}{c}{Case II}
& \multicolumn{2}{c}{Case III}  \\
& $\Lambda_{\rm EW}$ & $\Lambda_{\rm SS}$ & $\Lambda_{\rm EW}$ &
$\Lambda_{\rm SS}$
& $\Lambda_{\rm EW}$ & $\Lambda_{\rm SS}$ \\
\hline
$m^{}_1 ~({\rm eV} )$ & 0.2 & 0.24 & 0.2 & 0.24 & 0.2 & 0.24 \\
$\Delta m^2_{21} ~( 10^{-5} ~{\rm eV}^2 )$ & 8.0 & 22.15 & 8.0
& 17.59 & 8.0 & 22.40 \\
$\Delta m^2_{32} ~( 10^{-3} ~{\rm eV}^2 )$ & 2.5 & 3.66 & 2.5
& 3.68 & 2.5 & 3.66 \\
\hline
$\theta_{12}$ ($^\circ$) & 34 & 14.48 & 34
& 33.35 & 34 & 14.31 \\
$\theta_{23}$ ($^\circ$) & 45 & 42.48 & 45
& 43.98 & 45 & 43.52 \\
$\theta_{13}$ ($^\circ$) & 0 & $3.60 \times 10^{-2}$ & 0 &
1.12 & 0 & $2.77 \times 10^{-2}$ \\
\hline
$\delta$ ($^\circ$) & -- & 0 & -- & 81.41
& -- & 139.92 \\
$\alpha^{}_1$ ($^\circ$) & 0 & 0 & 260 & 273.63
& 80 & 80.13 \\
$\alpha^{}_2$ ($^\circ$) & 0 & 0 & 100 & 106.03
& 80 & 80.13 \\
\end{tabular}
\end{center}
\end{table}

\begin{table}
\caption{{\bf Near degeneracy} with $\Delta m^2_{32} < 0$:
numerical examples for radiative corrections to neutrino masses,
lepton flavor mixing angles and CP-violating phases from
$\Lambda_{\rm EW} \sim 10^2$ GeV to $\Lambda_{\rm SS} \sim
10^{14}$ GeV. The Higgs mass $m^{}_H = 140$ GeV (SM) or $\tan\beta
= 10$ (MSSM) has typically been input in our calculation.}
\begin{center}
\begin{tabular}{l|ll|ll|ll}
SM & \multicolumn{2}{c}{Case I} & \multicolumn{2}{c}{Case II}
& \multicolumn{2}{c}{Case III}  \\
& $\Lambda_{\rm EW}$ & $\Lambda_{\rm SS}$ & $\Lambda_{\rm EW}$ &
$\Lambda_{\rm SS}$
& $\Lambda_{\rm EW}$ & $\Lambda_{\rm SS}$ \\
\hline
$m^{}_1 ~({\rm eV} )$ & 0.2 & 0.28 & 0.2 & 0.28 & 0.2 & 0.28 \\
$\Delta m^2_{21} ~( 10^{-5} ~{\rm eV}^2 )$ & 8.0 & 15.28 & 8.0
& 15.27 & 8.0 & 15.28 \\
$\Delta m^2_{32} ~( 10^{-3} ~{\rm eV}^2 )$ & $-2.5$ & $-4.89$
& $-2.5$ & $-4.89$ & $-2.5$ & $-4.89$ \\
\hline
$\theta_{12}$ ($^\circ$) & 34 & 34.78 & 34
& 34.02 & 34 & 34.78 \\
$\theta_{23}$ ($^\circ$) & 45 & 44.95 & 45
& 44.98 & 45 & 44.97 \\
$\theta_{13}$ ($^\circ$) & 0 & $7.42 \times 10^{-4}$ & 0 &
$2.35 \times 10^{-2}$ & 0 & $5.68 \times 10^{-4}$ \\
\hline
$\delta$ ($^\circ$) & -- & 0 & -- & 89.55
& -- & 319.99 \\
$\alpha^{}_1$ ($^\circ$) & 0 & 0 & 260 & 259.60
& 80 & 80.00 \\
$\alpha^{}_2$ ($^\circ$) & 0 & 0 & 100 & 99.82 & 80 & 80.00 \\
\hline\hline
MSSM & \multicolumn{2}{c}{Case I} & \multicolumn{2}{c}{Case II}
& \multicolumn{2}{c}{Case III} \\
& $\Lambda_{\rm EW}$ & $\Lambda_{\rm SS}$ & $\Lambda_{\rm EW}$ &
$\Lambda_{\rm SS}$
& $\Lambda_{\rm EW}$ & $\Lambda_{\rm SS}$ \\
\hline
$m^{}_1 ~({\rm eV} )$ & 0.2 & 0.24 & 0.2 & 0.24 & 0.2 & 0.24 \\
$\Delta m^2_{21} ~( 10^{-5} ~{\rm eV}^2 )$ & 8.0 & 23.34
& 8.0 & 17.72 & 8.0 & 23.31 \\
$\Delta m^2_{32} ~( 10^{-3} ~{\rm eV}^2 )$ & $-2.5$ & $-3.61$
& $-2.5$ & $-3.58$ & $-2.5$ & $-3.61$ \\
\hline
$\theta_{12}$ ($^\circ$) & 34 & 13.69 & 34
& 33.45 & 34 & 13.85 \\
$\theta_{23}$ ($^\circ$) & 45 & 47.42 & 45
& 46.02 & 45 & 46.42 \\
$\theta_{13}$ ($^\circ$) & 0 & $3.70 \times 10^{-2}$ & 0 &
1.14 & 0 & $2.84 \times 10^{-2}$ \\
\hline
$\delta$ ($^\circ$) & -- & 0 & -- & 260.86
& -- & 139.93 \\
$\alpha^{}_1$ ($^\circ$) & 0 & 0 & 260 & 270.98
& 80 & 80.12 \\
$\alpha^{}_2$ ($^\circ$) & 0 & 0 & 100 & 104.88
& 80 & 80.12 \\
\end{tabular}
\end{center}
\end{table}

\end{document}